\documentclass[pra,reprint,twocolumn]{revtex4-1}

\usepackage{graphicx}

\usepackage{graphicx}
\usepackage{color}

\newcommand{\be}{\begin{equation}}
\newcommand{\ee}{\end{equation}}
\newcommand{\bea}{\begin{eqnarray}}
\newcommand{\eea}{\end{eqnarray}}

\begin{document}

\title{Quantum Dynamics in Noisy Backgrounds: \\ from sampling to dissipation and fluctuations} 

\author{O. Oliveira$^{1,2}$, W. de Paula$^{2}$, T. Frederico$^{2}$, M. S. Hussein$^{2,3}$}
\affiliation{$^1$CFisUC, Department of Physics, University of Coimbra, P-3004 516 Coimbra, Portugal \\
$^2$Instituto Tecnol\'ogico de Aeron\'autica, DCTA, 12228-900 S\~ao Jos\'e dos Campos, Brazil \\
$^3$Instituto de Estudos Avan\c{c}ados and Instituto de F\'{\i}sica, Universidade de S\~ao Paulo,
Caixa Postal 66318, 05314-970 S\~ao Paulo, SP, Brazil}

\date{\today}

\begin{abstract}
We investigate the dynamics of a quantum system coupled linearly to Gaussian white noise
using functional methods. 
By performing the integration over the noisy field in the evolution operator, we get an
equivalent non-Hermitian Hamiltonian, which evolves the quantum state 
with a dissipative dynamics. We also show that if the integration over the 
noisy field is done for the time evolution of the density matrix, a gain
contribution from the fluctuations, can be accessed in addition to the loss one 
from the non-hermitian Hamiltonian dynamics. 
We illustrate our study by computing analytically
the effective non-Hermitian Hamiltonian, which we found to be the complex 
frequency harmonic oscillator, with a known evolution operator. It leads 
to space and time localisation, a common feature of noisy quantum systems in general
applications. 
\end{abstract}


\maketitle

%=============================================================================
%=============================================================================
\section{Introduction and Motivation}

The interaction of quantum systems with a complex background can be simulated via the introduction of random fields $\xi (t,x)$.
The random fields account for residual interactions with the complex environment  or, when it is perturbed by an external field, 
can be associated with a noisy component in the external field as can happen, for example, in the interaction of a laser with electrons in atoms (see e.g.~\cite{Singh2007}).
An example for the first situation can be found in~\cite{Orth2013} where 
the spin-boson Hamiltonian is used to investigate the real-time dissipative dynamics of quantum impurities embedded in a macroscopic environment.
Relying on functional methods for quantum dissipative systems~\cite{Feynman1963}, the authors reformulated the original problem in terms of a stochastic Schr\"odinger
equation with a Gaussian noise coupled linearly to the quantum system.  
The spin-boson model is used across many areas of physics from quantum computing, to the investigation of dissipation-induced quantum phase transitions.

The consideration of noisy interactions to model complex systems is of  broad interest in science. Examples are provided by stochastic resonances (see e.g.~\cite{Wellens2004}),
which are relevant to many other fields as geology, engineer, biology and medicine.
Random fields are also a necessary ingredient to describe systems with random couplings as, for example, spin glasses or disordered
lattice systems~\cite{Fisher89,Katzer2015}.

Stochastic or random processes are invoked in many different situations
ranging from microscopic systems, as the investigation of the Dirac spectrum in quantum chromodynamics~\cite{Shuryak1993}, 
to large classical systems, as for example to describe turbulence in fluid dynamics~\cite{Benedict1996}.

In an application of a system driven by noise, one has to assume \textit{a priori} a given probability distribution for the random field. An usual choice is
to assume a Gaussian white noise.

Herein, we will focus on the dynamics of non-relativistic quantum systems coupled to random fields. We formulate the problem in configuration space
but, in principle, the method 
can be extended to any representation or applied to spin systems.
For a class of couplings between the quantum system and $\xi$ and for certain probability distributions,
the random fields can be integrated out exactly.

The fluctuations due to the random fields enable the definition of various types of averages where the combined role of the fluctuations and
dissipation can be investigated simultaneously. Although, we write all the necessary formalism to analyse the contributions coming from the fluctuations to
any correlation function, herein we focus mainly on the dissipative part.

The integration over the random fields 
for the evolution of the quantum state amplitude allows the identification of
an effective non-Hermitian Hamiltonian with negative imaginary part, ensuring that, for
sufficiently large time, the wave function $\psi (x,t)$ vanishes independently of the 
initial condition. The original Hamiltonian is recovered in the limit of vanishing noise.
The fluctuations can be accessed by computing the time evolution of the 
density matrix, which gives two parts, one associated with the dynamics provided 
by the dissipative non-Hermitian Hamiltonian, and another part associated with 
fluctuations of the random field. In this way, our formalism realizes  
the dissipation-fluctuation physics in the time evolution of the density matrix.

The emergence of non-Hermitian
effective Hamiltonian in the description of complex systems is a 
common occurrence in several fields. A well known example is nuclear collision theory, 
as described by the optical model \cite{Feshbach1, Feshbach2}. Thus the final results of 
our calculation as described below are supported by previous theories which were aimed at 
a simplified description of complex reactions.

Our method avoids large time computing simulations associated with many 
realisations necessary to build a statistical significant
ensemble to follow the 
dissipative aspect of the 
evolution of the quantum state. 
The method is sufficiently general  and can be applied in many 
different types of noisy quantum systems.

%==========================================================================
%==========================================================================
\section{Linear Coupling} 

Let us discuss the case where the random field has as a dipole type interaction with the quantum particle. As a
prototype one considers an electron in an hydrogen atom coupled to an intense linearly polarised laser field $F(t)$, perturbed by a stochastic force $\xi (t)$.
Such system was studied in~\cite{Singh2007,Singh2007A} relying on the following Hamiltonian (in atomic units $\hbar = m = e = 1$)
\begin{equation}
   H(p,x) = \frac{p^2}{2} + V(x)  + x \left\{ F(t) + \xi(t) \right\} \ ,
   \label{H:laser}
\end{equation}
where $V(x) = - 1/\sqrt{x^2 + a^2}$ is a non-singular Coulomb like potential.
Although, the following discussion uses the Hamiltonian (\ref{H:laser}), the conclusions are general for couplings of type $x \, \xi$ and are valid
beyond the one dimension and one particle quantum systems.

Let us consider a Gaussian white noise as in~\cite{Singh2007,Singh2007A}, i.e. the random variable satisfies the following relations
\begin{equation}
         \langle \xi(t) \rangle = 0 \qquad\mbox{ and }\qquad \langle \xi(t) \xi(t') \rangle = 2  D \, \delta (t - t') \ ,
\end{equation}
where the noise intensity $D$ is related to the variance of the Gaussian distribution $\sigma^2$.

The evolution described by the stochastic Hamiltonian (\ref{H:laser}) is nondeterministic and, therefore, to build a statistically meaningful solution of the time-dependent
stochastic Schr\"odinger equation, an average over many realisations of the system is required. For a given realisation, i.e. for a given $\xi (t)$, the evolution
of the system can be described by the propagator
\begin{equation}
    U_\xi ( x', t'; x, t) = \int \mathcal{D} q \mathcal{D}p  \exp\left\{ i \int^{t'}_t dt \bigg[ p \, \dot{q} - H(p,q) \bigg] \right\},
\end{equation}
where the integration is performed over the trajectories satisfying the boundary conditions $q(t') = x'$ and $q(t) = x$. A statistically meaningful
solution of the time dependent problem is given by summing over many $\xi$ fields distributed according to a Gaussian distribution. The
exact propagator reads
\begin{equation}
    U ( x', t'; x, t) = \int \, \mathcal{D} \xi \, \mathcal{D} q \, \mathcal{D}p ~ e^\Gamma \ ,
    \label{U}
\end{equation}
where $\mathcal{D} \xi$ stands for the continuum limit of $ \Pi_i \,  d \xi(t_i)$ and
\begin{equation}
 \Gamma = i \int^{t'}_t dt'' \, \bigg[ p \, \dot{q} - H(p,q) \bigg] - \frac{1}{2} \int^{t'}_t d t'' \, \frac{ \xi^2 (t'') }{ \sigma^2 ( t'') } \ .
 \label{Eq:gamma}
\end{equation}
The last term in (\ref{Eq:gamma}) represents a time dependent Gaussian probability distribution associated with the random variable $\xi (t)$,
which brings the average over the many realisations of the random field.

From the point of view of integration over $\xi$ in (\ref{U}), the integral is of Gaussian type and the exact integration gives
\begin{eqnarray}
  & &
     \int \, \mathcal{D} \xi ~ \exp\left\{  \int^{t'}_t dt'' \, \left[ - \frac{1}{2} \frac{ \xi^2 (t'') }{ \sigma^2 (t'')}  -  i \, q(t'') \,  \xi (t'') \right] \right\}  = \nonumber \\
   & &
   = \exp\left\{  - \frac{1}{2} \int^{t'}_t d t'' \, \sigma^2 ( t'' ) \,  q^2( t'' )  \right\} \ .
\end{eqnarray}
Then, the resulting evolution operator can be reinterpreted in terms of the new effective non-hermitian Hamiltonian
\begin{equation}
   H' (p,x,t) = \frac{p^2}{2} + V(x)  + x \, F(t)  - \frac{i}{2} \sigma^2(t) \, x^2 \ ,
    \label{Hfinal}
\end{equation}
which has only the degrees of freedom of the original quantum system, the deterministic external field $F(t)$ and a non-hermitian term
proportional to width of the Gaussian distributions. The variance $\sigma^2 (t)$ associated to the Gaussian distributions  is, from the point of
view of $H'$, an external field.

If one ignores the contribution of $V$ and $F$, the integration over the random fields gives rise to a pure imaginary oscillator
with negative imaginary frequency. Therefore, for a sufficient large time $\psi (x,t)$ vanishes independently of the initial condition.

For the particular case where V(x) is quadratic in x, F(t) = 0, and $\sigma (t)$ is time independent, the integration over the random field defines a 
damped complex harmonic oscillator.
The wave functions and energy spectrum of the complex harmonic oscillator, i.e. with a frequency $\omega = \omega_1 + i \omega_2$
with $\omega_1$, $\omega_2$ real and constants, were investigated in~\cite{Jannussis1986}.
For the general case,  the wave functions are no longer orthogonal but reduce to the usual harmonic oscillator functions in the limit $\omega_2 \rightarrow 0$.
Furthermore, one can define coherent states and for these states the Heisenberg uncertainty relation is verified
\begin{equation}
    \Delta x \, \Delta p = \frac{\hbar}{2} \, \sqrt{ 1 + \frac{\omega^2_2}{\omega^2_1}} > \frac{\hbar}{2} \ .
\end{equation}
The time-dependent eigenfunctions of the complex harmonic oscillator can be defined in $\mathcal{L}_2 (X ,T)$ and read
\begin{eqnarray}
\!\!\!
 \psi_n (x,t) & = & \left[ \frac{ 2 \, T \, \omega_2 \left( n + \frac{1}{2} \right)}{\exp\left(2 \, T \, \omega_2 \left( n + \frac{1}{2} \right)\right) - 1} \right]^{1/2}   \nonumber \\
 & &
 \exp\left\{ - i \,  t \, ( \omega_1 + i \omega_2 ) \left( n + \frac{1}{2} \right) \right\}  \psi_n (x,0)  .
\end{eqnarray}
For negative imaginary frequencies $\omega_2 < 0$ and for large times the wave function is driven to zero.

%%==========================================================
%%==========================================================
\section{Density Matrix: dissipation and fluctuations}

The density matrix for a quantum mechanical system is defined as
\begin{eqnarray}
   \rho (t; x_f, x_i) & = & \int dx ~dx' ~ \langle x_f | U(t) | x \rangle \, \langle x | \rho (0) | x' \rangle \nonumber \\
   & & \hspace{1.5cm}  \langle x' | U^\dagger (t) | x_i \rangle \ ,
   \label{Eq:rho_def}
\end{eqnarray}
where $\rho (0)$ is the density matrix at time $t = 0$ and $U(t)$ is the evolution operator.
The matrix elements of $U(t)$ can be written as functional integrals of type
\begin{equation}
  \langle x''  | U(t) | x' \rangle = \int \mathcal{D} x \, \mathcal{D} p ~ e^{ i \int^t_0 \left[ p \dot{q} - H(p,q) \right]} \ ,
\end{equation}  
where we have set $\hbar = 1$.
The functional integrals can be defined introducing a partition of the interval $[ 0 , t ]$, where $t_j = j \Delta t$ and
$j = 0, \dots , N$, and should be understood as the large $N$ limit of the multidimensional integral
\begin{widetext}
\begin{equation}
  \langle x''  | U(t) | x' \rangle  = \int \prod^{N-1}_{j=1} \frac{ dq_j \, dp_j}{2 \pi} \, \frac{ dp_0}{2 \pi} ~ 
  \exp\left\{ i \Delta t \sum^{N-1}_{n = 0} \left[ p_n \frac{q_{n+1} - q_{n}}{\Delta t} - H (p_n, q_n) \right] \right\}
\end{equation}  
where $q_j = q(t_j)$, $p_j = p(t_j)$, $q_0 = x'$ and $q_N = x$. The density matrix can also be put into a functional
form as follows
\begin{eqnarray}
   \rho (t; x_f, x_i) & = & \int dx ~dx' ~ \langle x_f | U(t) | x \rangle \, \langle x | \rho (0) | x' \rangle \,  \langle x_i | U (t) | x' \rangle^* \nonumber \\
   & = & \int dq_0 \, dQ_0 ~ \left[ \prod^{N-1}_{j=1} \frac{ dq_j \, dp_j}{2 \pi} \, \frac{ dp_0}{2 \pi}\right] ~ \left[ \prod^{N-1}_{j=1} \frac{ dQ_j \, dP_j}{2 \pi} \, \frac{ dP_0}{2 \pi} \right]
   ~           \rho (t=0; q_0, Q_0) 
    \nonumber \\
    & & \hspace{2cm} \exp\left\{ i \, \Delta t \sum^{N-1}_{n = 0} \left[  p_n \frac{q_{n+1} - q_n }{\Delta t} - H(p_n,q_n) - P_n \frac{Q_{n+1} - Q_n }{\Delta t} + H(P_n,Q_n) \right] \right\}
    \nonumber \\
   & = & \int  \left[ \prod^{N-1}_{j=0} \frac{ dq_j \, dp_j}{2 \pi} \right] ~ \left[ \prod^{N-1}_{j=0} \frac{ dQ_j \, dP_j}{2 \pi}  \right]
   ~           \rho (t=0; q_0, Q_0) 
    \nonumber \\
    & & \hspace{2cm} \exp\left\{ i \, \Delta t \sum^{N-1}_{n = 0} \left[  p_n \frac{q_{n+1} - q_n }{\Delta t} - H(p_n,q_n) - P_n \frac{Q_{n+1} - Q_n }{\Delta t} + H(P_n,Q_n) \right] \right\}
\end{eqnarray}
where $(q,p)$ refer to the canonical variables associated to the partition of $\langle x_f | U(t) | x \rangle$ and $(Q,P)$
are the canonical variables associated to $\langle x_i | U(t) | x' \rangle$.

Let $H_0$ be the Hamiltonian of a noiseless quantum system which couples linearly to the white noise $\xi (t)$. The total Hamiltonian of system reads
$H = H_0 + \lambda \, x \, \xi(t)$, where $\lambda$ is a real coupling constant setting the strength of the coupling to the random field. For a given realisation 
of the noise $\xi (t)$ one can associate the density matrix
\begin{eqnarray}
   \rho_\xi (t; x_f, x_i) & = &  \int  \left[ \prod^{N-1}_{j=0} \frac{ dq_j \, dp_j}{2 \pi} \right] ~ \left[ \prod^{N-1}_{j=0} \frac{ dQ_j \, dP_j}{2 \pi}  \right]
   ~           \rho (t=0; q_0, Q_0) 
    \nonumber \\
    & & 
    \exp\left\{ i \, \Delta t \sum^{N-1}_{n = 0} \left[  p_n \frac{q_{n+1} - q_n }{\Delta t} - H_0(p_n,q_n) - P_n \frac{Q_{n+1} - Q_n }{\Delta t} + H_0(P_n,Q_n) - 
                        \lambda \left( q_n - Q_n \right) \, \xi_n \right] \right\} \ .
\end{eqnarray}
The averaged density matrix is  obtained by integrating $\rho_\xi$ over the variabes $\xi$ which follow a gaussian distribution and, therefore, is given by
\begin{eqnarray}
   \overline\rho (t; x_f, x_i) & = &  \int \left[ \prod^{N-1}_{j = 0} \frac{d \xi_j}{\sqrt{ 2 \pi \sigma^2_j / \Delta t}} \right] ~ \left[ \prod^{N-1}_{j=0} \frac{ dq_j \, dp_j}{2 \pi} \right] ~ \left[ \prod^{N-1}_{j=0} \frac{ dQ_j \, dP_j}{2 \pi}  \right]
   ~           \rho (t=0; q_0, Q_0) 
    \nonumber \\
    & & 
    \exp\left\{ i \, \Delta t \sum^{N-1}_{n = 0} \left[  p_n \frac{q_{n+1} - q_n }{\Delta t} - H_0(p_n,q_n) - P_n \frac{Q_{n+1} - Q_n }{\Delta t} + H_0(P_n,Q_n) - 
                        \lambda \left( q_n - Q_n \right) \, \xi_n \right] \right\} \nonumber \\
    & &
    \exp\left\{ - \frac{1}{2}  \Delta t \sum^{N-1}_{n = 0} \frac{\xi_n}{\sigma^2_n}  \right\} \ .
\end{eqnarray}
The integrals of the white noise are gaussian integrals which can be performed exactly given the following averaged density matrix
\begin{eqnarray}
   \overline\rho (t; x_f, x_i) & = &  \int  \left[ \prod^{N-1}_{j=0} \frac{ dq_j \, dp_j}{2 \pi} \right] ~ \left[ \prod^{N-1}_{j=0} \frac{ dQ_j \, dP_j}{2 \pi}  \right]
   ~           \rho (t=0; q_0, Q_0) 
    \nonumber \\
    & & 
    \exp\left\{ i \, \Delta t \sum^{N-1}_{n = 0} \left[  p_n \frac{q_{n+1} - q_n }{\Delta t} - H_0(p_n,q_n) - P_n \frac{Q_{n+1} - Q_n }{\Delta t} + H_0(P_n,Q_n)  \right] \right\} \nonumber \\
    & & 
    \exp\left\{ - \frac{\lambda^2}{2}  \Delta t \sum^{N-1}_{n = 0} \sigma^2_n \, \left( q_n - Q_n \right)^2  \right\} \nonumber \\
    & = &     \int  \left[ \prod^{N-1}_{j=0} \frac{ dq_j \, dp_j}{2 \pi} \right] ~ \left[ \prod^{N-1}_{j=0} \frac{ dQ_j \, dP_j}{2 \pi}  \right]
   ~           \rho (t=0; q_0, Q_0) 
    \nonumber \\
    & & 
    \exp\left\{ i \, \Delta t \sum^{N-1}_{n = 0} \left[  p_n \frac{q_{n+1} - q_n }{\Delta t} - H_{eff}(p_n,q_n) - P_n \frac{Q_{n+1} - Q_n }{\Delta t} + H^*_{eff}(P_n,Q_n)  \right] \right\} \nonumber \\
    & & 
    \exp\left\{  \lambda^2  \Delta t \sum^{N-1}_{n = 0} \sigma^2_n \,  q_n \,  Q_n  \right\} 
    \label{Eq:ave_rho}
\end{eqnarray}
\end{widetext}
where the new effective non-hermitian dissipative Hamiltonian reads
\begin{equation}
  H_{eff}(p,q) = H_0(p,q) - \frac{i}{2} \, \sigma^2 \,\lambda^2 q^2 \ .
\end{equation}  
If one ignores the last term in (\ref{Eq:ave_rho}), the time evolution of $\overline\rho$ requires only the knowledge of $H_{eff}$ and the quantum system 
is dissipative. If the dynamics associated with $H_{eff}$ favours the collapse of the quantum system, the last term in  (\ref{Eq:ave_rho}) introduces correlations 
between the paths $x \rightarrow x_f$ and $x' \rightarrow x_i$ which favours the revival of the quantum system. 

Collapses and revivals of quantum systems are familiar phenomena in many areas of quantum physics as e.g. quantum optics. 
The Jaynes-Cummings model~\cite{JaynesCummings63,Rempe1987}
is a well known model where successive collapses and revivals are present and whose interest goes beyond
quantum optics. The Jaynes-Cummings model is related to Caldeira--Leggett model~\cite{Caldeira1981}, a popular quantum mechanical  system set to
include dissipation in a system coupled to a heat bath.

In order to build a solution for $\overline\rho$ one take the large $N$ limit of (\ref{Eq:ave_rho}) and write the above expression as a functional
integral
\begin{widetext}
\begin{eqnarray}
   \overline\rho (t; x_f, x_i)  &  = &   \int  \!\! \mathcal{D}q \, \mathcal{D} p \, \mathcal{D} Q \, \mathcal{D} P \,    \rho \big(q (0), Q (0) \big) \,
    \exp\left\{ i \, \int^t_0 dt ~\left[  p \dot{q} - H_{eff}(p,q) - P \dot{Q}  + H^*_{eff}(P,Q) -  i \, \lambda^2  \sigma^2 \,  q \,  Q \right]  \right\}  
    \label{Eq:rho_ave_functional}
\end{eqnarray}
\end{widetext}
with the following boundary conditions $q(t) = x_f$ and $Q(t) = x_i$.

An important quantity to consider which would clearly exhibit the dissipation - fluctuation aspect of the $\xi$-averaged density matrix  is its time derivative. For conservative system governed by an Hermitian Hamiltonian, $H_0$, this equation is the so-called Pauli evolution equation,
\begin{equation}
i\dot\rho_0 = \big[H,\rho_0\big] \ .
\end{equation}
Adding the white noise and using the density matrix for the full Hamiltonian $H_{\xi} = H_0 + \lambda \, \xi x$, the equation for the corresponding density matrix, $\rho_{\xi}$ still satisfies the same Pauli equation,
\begin{equation}
i\dot\rho_{\xi} = \big[H_{\xi}, \rho_{\xi}\big] \ .
\end{equation}
On the other hand the equation for the time derivative of the $\xi$-averaged density matrix, would contain a dissipative term (loss term) and a 
fluctuation term (correlation, gain, term), viz,
\begin{equation}
i \dot{\overline{\rho}} = \big[H_{0}, \overline{\rho}\big] - L \overline{\rho} + G\overline{\rho}
\end{equation}
where $L\overline{\rho}$, the loss term, and $G\overline{\rho}$, the gain term, are short hand notations for the action of the dissipation term, 
$-i \lambda^2 \sigma^{2}q^2/2$,   and the correlation term,  $ i \lambda^2  \sigma^2 \,  q \,  Q$, respectively, see Eq.~(\ref{Eq:rho_ave_functional}).  In fact, the loss term is just,
\begin{equation}
L\overline{\rho} = \frac{1}{2}\lambda^2 \sigma^2 \big[ q^2, \overline{\rho}\big] \ .
\end{equation}
The evolution equation of $\overline{\rho}$ above is an important formal entity which exhibits in a transparent way the dissipation-fluctuation aspect of the action of the white noise 
on the physical system. Further, it supplies a mean to find the evolution of the averages of physical, observable quantities. 

A way to actually calculate the gain term above, is to resort to the source method.
 Introducing the sources $j$ and $J$ which couple to $q$ and $Q$, respectively, one can write
\begin{equation}
  \overline\rho (t; x_f, x_i)  =  \left. \exp\bigg\{ i \, S' \left[   i \frac{\delta}{\delta j} \, , \, - i  \frac{\delta}{\delta J}\right]  \bigg\} ~ Z_\rho[j,J] \right|_{j=J=0}
  \label{Eq:rho_evolution}
\end{equation}
where
\begin{widetext}
\begin{equation}
S' \left[   q \, , \, Q \right] = \int^t_0 dt ~\left[  -  i \, \lambda^2  \sigma^2 \,  q \,  Q \right]
\end{equation}
and
\begin{equation}
   Z_\rho[j,J]    =      \int  \mathcal{D}q \, \mathcal{D} p \, \mathcal{D} Q \, \mathcal{D} P ~    \rho \big(q (0), Q (0) \big) ~
    \exp\left\{ i \, \int^t_0 dt ~\left[  p \dot{q} - H_{eff}(p,q) - P \dot{Q}  + H_{eff}^{\dagger}(P,Q)  -  j q  + J  Q \right]  \right\}  \ .
\end{equation}
\end{widetext}
Expression (\ref{Eq:rho_evolution}) provides the formal solution to compute the average value of density matrix. For small enough variances,
$\exp\{ i \, S' \left[   i  \delta/\delta j ,   -i  \delta/\delta j \right]\}$ can be expanded in powers of $\sigma^2$ and the correlations between the paths
can be written as a power series of the variance of the white noise.

%================================================================================
%================================================================================
\section{Example: Noisy Harmonic Oscillator} 

Let us consider the case of an harmonic oscillator defined by the potential $V(x) = m \, \omega^2  \, x^2/ 2$ in interaction with
a Gaussian white noise described by a term like $x \, \xi$ in the Hamiltonian.  Following the prescription described above, %in the previous section,
after the functional integration over $\xi$, the new effective Hamiltonian is 
\begin{equation}
   H' = \frac{p^2}{2 \, m} + \frac{1}{2} m \, \omega^2 \, x^2 - \frac{i}{2} \sigma^2(t) x^2 \ .
\end{equation}
Setting $\sigma^2 (t) = m \, \overline\sigma^2 > 0$ and introducing the complex frequency squared $\Omega^2 =  \omega^2 - i \, \overline\sigma^2$,
the propagator for the noisy harmonic oscillator and a time independent variance is given by (see e.g.~\cite{Itzykson80})
\begin{widetext}
\begin{equation}
U (x',t';x,t)  = 
\left[ \frac{ m \, \Omega \, e^{-i \pi / 2} }{2 \pi \, \sin ( \Omega (t' - t) ) } \right]^{1/2} 
  \exp\left\{ \frac{i \, m \, \Omega}{2} \left[ (x'^{\, 2} + x^2 ) \cot ( \Omega (t' - t) ) - \frac{ 2 \, x' \,  x}{\sin ( \Omega (t' - t) )}\right] \right\} \ .
\end{equation}
Writing $\Omega = \omega_1 - i \, \omega_2$, for a particle created at time $t = 0$ and at position $x = 0$  the wave function for positive time is
\begin{eqnarray}
  \psi (x, t)  =  U (x,t;0,0) & = & \left[ \frac{ m \, \Omega \, e^{-i \pi / 2} }{2 \pi \, \sin ( \Omega t ) } \right]^{1/2} \,
                                                \exp\left\{ \frac{i \, m \,  x^2 \, \Omega}{2} \, \cot ( \Omega t) \right\} \nonumber \\
  & = & \left[ \frac{ m \, \left( \omega_1 - i \, \omega_2 \right) \, e^{-i \pi / 2} }{2 \pi \, \left[ \sin ( \omega_1 t ) \cosh( \omega_2 t) - i \, \cos ( \omega_1 t ) \sinh( \omega_2 t) \right] }
            \right]^{1/2}  \nonumber \\
 & & \qquad
  \exp\left\{ \frac{ m \,  x^2 }{4} \, \frac{\omega_2\sin(2 \omega_1 t) - \omega_1\sinh(2 \omega_2 t) + i \left[ \omega_2 \sinh( 2 \omega_2 t ) + \omega_1 \sin(2 \omega_1 t) \right]}{\sin^2( \omega_1 t) \cosh^2(\omega_2 t) + \cos^2( \omega_1 t) \sinh^2( \omega_2 t)} \right\}  \ .
  \label{psi:oscilador}
\end{eqnarray}
\end{widetext}
In the small width limit such that $\overline\sigma^2 / \omega^2 \ll 1$, $\omega_1 \approx \omega$ and
$\omega_2 \approx \overline\sigma^2 /  2 \omega > 0$. In the following, we will always assume $\omega_1 , \omega_2 \geq 0$.

For large  $t$, the wave function is exponentially damped both in space and in time directions
\begin{eqnarray}
  \psi (x, t)  &   \approx &
   \sqrt{ \frac{ m \, \left( \omega_1 - i \, \omega_2 \right)  }{\pi } }   ~ \exp\left\{- i \frac{\omega_1 t}{2} \right\} \, \exp\left\{- \frac{\omega_2 t}{2} \right\}
   \nonumber \\
   & &
   \qquad \exp\left\{ \frac{ m \,  x^2 }{2} \,  \left( - \omega_1 + i \omega_2 \right) \right\}   \ .
  \label{psi:oscilador:large}
\end{eqnarray}
This means that the quantum system is localised in space and has a finite time life.
The mean lifetime of the asymptotic noisy quantum oscillator $\tau = 2 / \omega_2$ is controlled by the imaginary part of the complex frequency $\Omega$.
In the small width limit as defined above $\tau = 4 \omega / \overline\sigma^2$ and therefore smaller Gaussian widths imply longer mean lifetimes. Indeed, in the
limit where $\overline\sigma^2 \rightarrow 0$, the mean lifetime becomes infinite. On the other hand, the space localisation of the noisy quantum oscillator
is controlled by the real part of the complex frequency $\Omega$. One can define the penetration depth $\lambda = \sqrt{2/ m \omega_1}$ which becomes
$\lambda = \sqrt{2/ m \omega}$ in the small width limit.

\begin{figure*}[t] %  figure placement: here, top, bottom, or page
   \centering
   \includegraphics[scale=0.29]{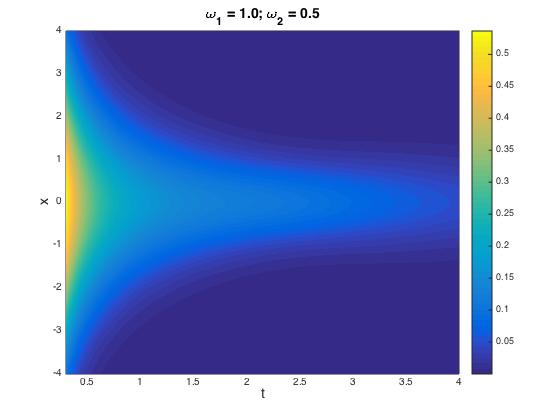}
   \includegraphics[scale=0.29]{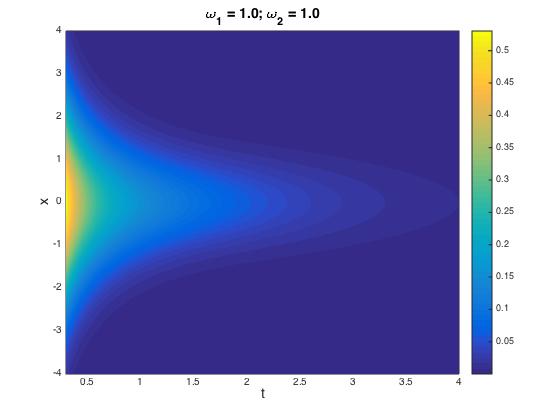}
   \includegraphics[scale=0.29]{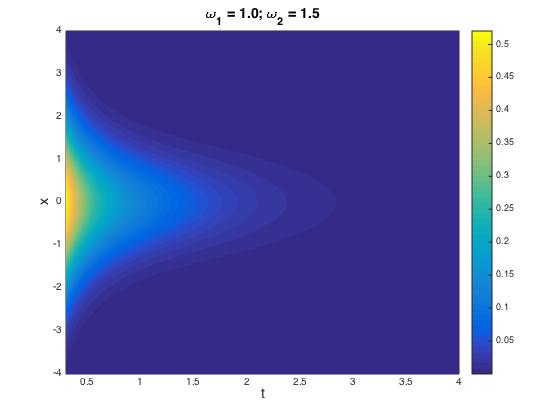}
   \caption{$|\psi (x,t)|^2$ for the noisy quantum oscillator and for various complex frequencies $\Omega = \omega_1 - i \omega_2$.}
   \label{fig:wavefunction}
\end{figure*}

The localisation in time and in space are general characteristics of the noisy quantum oscillator that occur even when $\omega$ vanishes, i.e. for a pure
noisy oscillator. In such case $\omega_ 1 = \omega_2 =\overline\sigma /  \sqrt{2} $ with $\tau$ and $\lambda$ being a measure of
the width of the Gaussian noise. It follows, that a very short (long)  lifetime implies that the system is spread over a very small (large) space region.
Surprisingly, a dipole like coupling of a quantum system with a Gaussian noise confines the quantum system.

From the general expression (\ref{psi:oscilador}) one concludes that if the localisation in the time direction is a general property of the noisy quantum
oscillator, localisation in the space direction can only occurs for times such that
\begin{equation}
 \omega_2\sin(2 \omega_1 t) - \omega_1\sinh(2 \omega_2 t) < 0 \ .
\end{equation}
If this condition is not fullfiled, the system can spread over the entire space.
From the above expression one can conclude that space localisation always occur for sufficient large times.
For the special case where $\omega_1 = \omega_2$, as in a pure noisy quantum oscillator,
the inequality is satisfied for all $t$ and the pure noisy quantum oscillator is a localised system.

\begin{figure}[t] %  figure placement: here, top, bottom, or page
   \centering
   \includegraphics[scale=0.2]{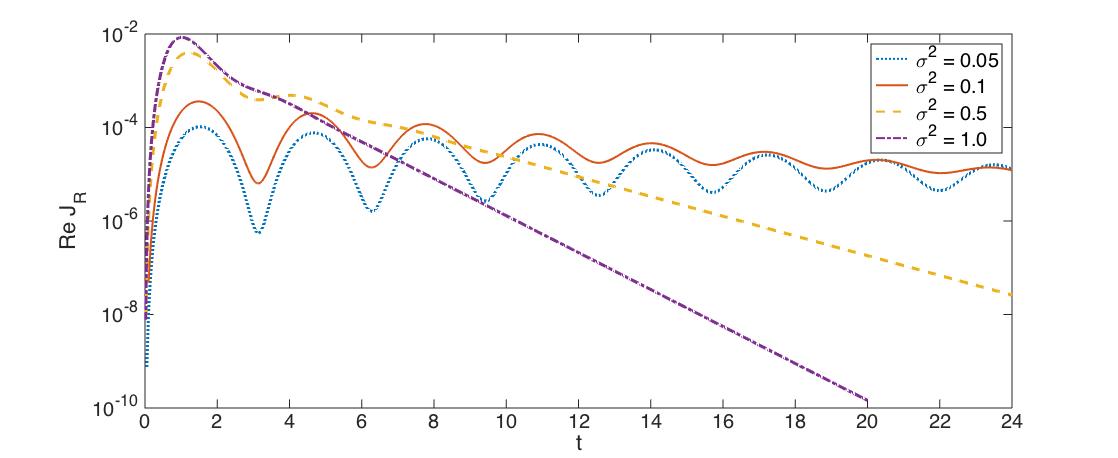}
   \caption{$J_R (x,t)$ computed after propagating the ground state of the harmonic oscillator from $t = 0$. For sufficiently large times, in all cases $J_R$ is
                  exponentially damped. See text for details.}
   \label{fig:corrente}
\end{figure}

The square of the wave function $|\psi (x,t)|^2$ for a noisy quantum oscillator is reported in Fig.~\ref{fig:wavefunction}
for three different complex frequencies $\Omega$. As discussed previously, the oscillator is localised in space and its wave function vanishes for sufficient large times,
even for a pure noisy quantum oscillator represented by $\omega_1 = \omega_2 = 1$ in the figure.

In Fig.~\ref{fig:corrente} we show the current defined by
\begin{equation}
   J_R (x,t) = \Re \left[ -i \, \psi^* (x,t) \, \frac{\partial \psi (x,t)}{\partial x} \right]
\end{equation}
computed at $x = 0.5$ (arbitrary units) after propagating the ground state of an harmonic oscillator with $m = 1$ and $\omega = 1$ from $t = 0$
and for various $\sigma^2$. In all cases one can identify an oscillatory behaviour for small and medium $t$ values followed by an exponential damping of
the system for sufficient large times. The pattern observed in Fig.~\ref{fig:corrente} follows closely that observed in similar figures for $J_R$
for the cases reported in~\cite{Singh2007}, where an electron in an hydrogen atom perturbed by a noisy laser beam was investigated.

It is instructive to discuss the spectrum of the complex oscillator, as it represents a variety of physical phenomena, such as the multi-phonon giant resonances in nuclei 
\cite{hussein1} and also atomic clusters \cite{hussein2}, as well as barrier tunneling influenced by coupling to a damped oscillator \cite{Takigawa}. 
One dimensional oscillator has the usual harmonic spectrum
$E_{n} = (n + 1/2)\hbar \omega$.
If $\omega$ is complex with a negative imaginary part, as our model exhibits, then $\Omega = \omega_1 -i\omega_2$. Identifying the width of a resonant state
by $\Gamma_n = (2n + 1)\hbar \omega_2$, then the spectrum becomes
$E_n = (n + 1/2)\hbar \omega_1 - i \, \Gamma_n/2$,
showing that the states of the oscillator become resonances with lifetimes $\tau_n = \hbar/ \Gamma_n$. 
To be physically consistent, the ground state of the oscillator is stable for sufficiently small noise levels as can be seen in Fig.~\ref{fig:corrente}, and 
thus the factor 1 in (2n + 1) in $\Gamma_{n}$ must be ignored. The widths of the first few 
states are thus, given in terms of the widths of the corresponding one phonon, two phonon, 
three phonon states etc. 
$\Gamma_1 = 2\hbar \omega_2, \Gamma_2 = 2\Gamma_1, \Gamma_3 = 3\Gamma_1, $ etc. Such relations 
are approximate as these phonons are bosons and accordingly the wave function of n-phonon 
states must be symmetric and contain a factor $1/\sqrt{n!}$. This factor would modify the 
expressions for the widths, as discussed in \cite{hussein1}. 

For the case where the coupling to the random field is strong and the perturbative approach is no longer applicable, 
which should be the case in many physical applications, 
then the ground state of the system is no longer stable and decays exponentially. This type of situations is illustrated in Fig.~\ref{fig:corrente}
setting $\omega_2 \sim \omega_1$ which configures a case where
the absorptive effective interaction competes with the harmonic oscillator potential.

\subsection{Fluctuations and Dissipation}

For an harmonic oscillator coupled linearly to a white noise, the effective non-hermitian Hamiltonian is given by 
$H_{eff} = p^2/2m + m \Omega^2 x^2 /2$, where $\Omega^2 = \omega^2 - i \lambda^2\sigma^2$ with $\omega$ being the natural frequency of the oscillator.
The time derivative of $\overline\rho$ can be computed from (\ref{Eq:ave_rho}) by looking at the difference $\overline\rho ( t + \Delta t ) - \overline\rho (t)$.
After some straightforward algebra it comes that
\begin{widetext}
\begin{eqnarray}
  i  \frac{d \overline\rho (t)}{dt} & = & \int \left[ \prod^{N-1}_{j=0} \frac{ dq_j \, dp_j}{2 \pi} \right] ~ \left[ \prod^{N-1}_{j=0} \frac{ dQ_j \, dP_j}{2 \pi}  \right]
                   \overline\rho (q_0, Q_0)  \left[H_{eff} (p_{N-1}, x_f ) - H^*_{eff} (P_{N-1}, x_i) + i \, m \, \lambda^2 \sigma^2 x_f x_i\right] \nonumber \\
   & = & -\, F[ \overline\rho ] - \frac{m}{2} \bigg[  \omega^2 (x^2_i - x^2_f) + i \,  \lambda^2 \sigma^2 (x_i - x_f)^2 \bigg] \, \overline\rho (t)
   \label{EQ:time_evo_rho}
\end{eqnarray}
where
\begin{eqnarray}
F[ \overline\rho ] & = &
 \int \left[ \prod^{N-1}_{j=0} \frac{ dq_j \, dp_j}{2 \pi} \right] ~ \left[ \prod^{N-1}_{j=0} \frac{ dQ_j \, dP_j}{2 \pi}  \right]
                   \overline\rho ( q_0, Q_0)  \frac{ P^2_{N-1} - p^2_{N-1}}{2 \, m} \nonumber \\
                   & &
                   \exp\left\{ i \Delta t \sum^{N-1}_{n = 0} 
                   \left[ p_n \frac{q_{n+1} - q_n}{\Delta t} - H_{eff} (p_n,q_n) - P_n \frac{Q_{n+1} - Q_n}{\Delta t} + H^*_{eff} (P_n,Q_n)  - i m \lambda^2 q_nQ_n\right] \right\}
                   \ .
\end{eqnarray}
\end{widetext}
It follows from the differential equation~(\ref{EQ:time_evo_rho}) that if the contribution of $F[ \overline\rho]$ is subleading, then the system collapses for sufficiently large
times. On the other hand, in $F$ the integration over $P_{N-1}$ and $p_{N-1}$ introduces a term which behaves like
\begin{displaymath}
 \left( \frac{2m}{\Delta t}\right)^2 \left[ (x_i - Q_{N-1})^2 - (x_f - q_{N-1})^2  - i \, \Delta t \right]  \times \mbox{ integral }
\end{displaymath}
and, depending on the relative values of $x_i$ and $x_f$, can be translated into the differential equation in a gain, i.e. $\overline\rho$ grows with $t$, or a loss of the system
with $\overline\rho$ vanishing for sufficiently large times.

%==================================================================================
%==================================================================================
\section{Discussion and Summary} 

In the present work we  investigated the coupling of a quantum system with a Gaussian white noise.
For a dipole coupling, i.e. proportional to the position of the quantum system, the noise can be integrated explicitly using functional methods.
In this way, one avoids the time consuming construction of statistical significant ensembles to follow the evolution of the wave function.

The integration over the Gaussian white noise introduces a new average which can be performed before or after the integration over the dynamical
variables. Then, for any correlation function $C(t_1, \cdots, t_n) =  \langle x_f | q(t_1) \cdots q (t_n) | x_i \rangle$ the integration over the random field can be perform either after
computing $C(t_1, \cdots, t_n) $ for a given realisation of the $\xi (t)$, i.e.
\begin{equation}
    C(t_1, \cdots, t_n) = \int \mathcal{D} \xi ~ C_\xi(t_1, \cdots, t_n) ~ e^{ -\frac{1}{2} \int^{t_f}_{t_i} \frac{\xi^2 (t)}{\sigma^2 (t)}}
\end{equation}
where $C_\xi$ stands for the expectation value $\langle x_f | q(t_1) \cdots q (t_n) | x_i \rangle$ for a given $\xi (t)$ field, or perform first the integration over the
random fields and define an effective non-hermitian Hamiltonian to compute $C(t_1, \cdots, t_n)$ in terms of $H'$. 
The interplay between the two different kinds of averages allow to describe the fluctuation-dissipation characteristics that are typically associated with a
complex quantum system.

For the latter type of averages,
the integration replaces the original Hamiltonian by a new non-hermitian effective Hamiltonian $H'$. As argued before and shown explicitly for the noisy
quantum oscillator, the dynamics associated to $H'$ is dissipative and localised, i.e. $\psi (x,t)$ vanishes for sufficient large times or for
sufficient large $|x|$.

This exponential damping in time and space is a general property associated with the non-hermitian Hamiltonian $H'$. For example in~\cite{Orth2013},
by solving an equivalent stochastic Schr\"odinger equation, the authors observed an exponential decay in the long time behaviour of the spin correlation functions.

Although we have considered a particular type of coupling between a quantum system and the Gaussian white noise, the procedure can be generalised beyond
the dipole coupling. For example, for  $H = H_0 + g \, \xi$, where $g$ represents a general operator, the integration considered here
generates the following effective Hamiltonian $H' = H_0 - i \, \sigma^2 \, g^2 / 2$. This class of Hamiltonian includes, for example, the case of the
nonlinear Schr\"odinger equation subject to random noise used to describe the evolution of  Bose-Einstein condensates, i.e. the Gross--Pitaevskii equation,
to investigate nonlinear photonics,  Langmuir waves in plasmas among other systems -- see e.g.~\cite{Chen2009,Cardoso2010} and references there in.
Furthermore, besides the linear coupling in $\xi$, the functional integration can also be performed exactly up to quadratic terms
$g_1 \, \xi + g_2 \, \xi^2 / 2$, where $g_1$ and $g_2$ represent possible quantum operators. In this case, the new effective Hamiltonian reads
$H' = H_0 + i \, g_1 \, ( g_2 - 1/\sigma^2)^{-1} g_1$.

As a final remark, we would like to show that the above procedure can be extended to include many-body interactions in a particle independent approach to a system
of identical particles.
For a system of identical particles one should
consider a single white noise which should be coupled to $\sum_i x_i$. The total Hamiltonian reads
\begin{equation}
   H = H_0 + \left( \sum_i x_i \right) \xi (t) \ ,
\end{equation}
where $H_0$ is the Hamiltonian for the many-particle system with zero noise. It follows that the new effective Hamiltonian is given by
\begin{equation}
   H' = H_0 - \frac{i}{2} \sigma^2  \left( \sum_i x_i \right)^2
\end{equation}
with the new non-hermitian interaction being proportional to the square of the position of 
the center of mass of the identical particle system. This non-hermitian operator
is responsible for the damping in time of the many-body system and also by its 
localisation in space. The non-hermitian interaction, 
 namely the dissipative force, drives the  many-particle system 
to collapse to its center of mass position for sufficiently large times, however  
the contribution from the fluctuation has to be accounted, and it will soften 
this sharp behaviour with a gain contribution as opposed to the losses 
from dissipation.

In summary, we propose a general method to describe the dynamics of quantum systems coupled to a Gaussian white noise
and that takes into account the dissipative and fluctuations aspects of a noisy quantum system. 
The dissipative aspect of the quantum evolution can be found by 
integrating directly the random fields in the definition of the partition 
function of the theory. This procedure establishes a mapping between 
quantum systems driven by white noise and non-hermitian dissipative Hamiltonians,
associated with losses, namely probability flows from the explicit degrees of freedom to the implicit ones
simulated by the noisy field. 
The exact integration and the use of $H'$ avoid the statistical sampling to compute the evolution of the  amplitude of 
the  quantum state. 
On the other hand, 
the  contributions from the fluctuations can be accessed 
by computing the time evolution of the density matrix and averaging over 
the stochastic field after. 
That gives two terms to the density evolution: one 
follows from the non-hermitian Hamiltonian dynamics with dissipation and corresponding 
losses and another one associated with a gain contribution from 
the fluctuations induced by the coupling of the quantum system to the white noise.

The method was formulated in configuration space but, in principle, it can be translated 
into any other representation or applied to spin systems
as those used in quantum computing.

For a dipole coupling to the white noise, the effective Hamiltonian is the complex 
frequency harmonic oscillator, with an analytical evolution operator.
It leads to space and time localisation, a common feature of noisy quantum systems.
In this case, the current reproduces the same pattern as observed when one uses stochastic 
methods~\cite{Singh2007}. Furthermore, the complex frequency naturally 
gives a width to the oscillator states which become resonant states. 

The effective quantum Hamiltonian $H'$ can be used to describe several many-body  
physical phenomena, such 
as Stochastic Resonances and Bose-Einstein Condensation, and, in particular, the complex  
frequency harmonic
oscillator can provide the starting point for a more general approach to the dynamics of 
such type of systems under the stochastic dipole interaction. So far only the losses are 
accounted explicitly by our example. To complete the picture the 
gain contribution to the density matrix should be computed, that task is 
beyond the present work and  we leave it for a future investigation.

%%==========================================================
%%==========================================================
\section*{Acknowledgements}

The authors acknowledge financial support from the Brazilian
agencies FAPESP (Funda\c c\~ao de Amparo \`a Pesquisa do Estado de
S\~ao Paulo) and CNPq (Conselho Nacional de Desenvolvimento
Cient\'ifico e Tecnol\'ogico). OO acknowledges financial support from grant 2014/08388-0 from S\~ao Paulo Research Foundation (FAPESP).
MSH acknowledges support from CAPES/ITA-PVS Fellowship program, and FAPESP/CEPID-CEPOF.

%%===============================================================
%%===============================================================

%=======================================================================
%=======================================================================

\end{document}